\documentclass[aps,prl,twocolumn,groupedaddress]{revtex4}
\usepackage{amssymb}
\usepackage{amsmath}
\usepackage{cases}
\usepackage{graphicx}

\begin{document}

\title{The Sands of Time Run Faster Near the End}

\author{Juha Koivisto and Douglas J. Durian}
\affiliation{Department of Physics and Astronomy, University of Pennsylvania, Philadelphia, Pennsylvania 19104-6396, USA }

\date{\today}

\begin{abstract}
Submerged granular hoppers exhibit an unexpected surge in discharge rate as they empty [Wilson {\it et al.} 2014].  With a more sensitive apparatus, we find that this surge depends on hopper diameter and also happens in air --- though the effect is smaller and previously unnoticed.  We also find that the surge may be turned off by fixing the rate of fluid flow through the granular packing.  With no flow control, dye injected on top of the packing gets drawn into the grains, at a rate that increases as the hopper empties.  Thus we conclude that the surge is caused by a self-generated pumping of fluid through the packing.  We successfully model this effect via a driving pressure set by the exit speed of the grains.  This highlights a surprising and unrecognized role that interstitial fluid plays in setting the discharge rate, and likely also in controlling clog formation, for granular hoppers whether in air or under water.
\end{abstract}

\maketitle

Hourglasses are filled with sand, rather than water, because the discharge rate of the grains is constant.  In particular it does not decrease as the filling height of the material in the upper chamber goes down, as it would for water.  This feature, and the variation of discharge rate with grain and orifice size, is captured empirically by the ``Beverloo equation'' \cite{Beverloo, NeddermanSavage}; however, the fundamental explanation is still under active research \cite{MazaGM07, AguirrePRL2010, Hannah_GM10, JandaPRL12, RubioLargoPRL15, DunatungaJFM15}.  The difficulty is that the grains have both solid-like and liquid-like behavior near the orifice.  Intuitively, the discharge speed of the grains is set by ephemeral arches and free-fall though a distance equal to the orifice size.  Analogous behavior was recently found for submerged granular hoppers, where the Beverloo equation was generalized by using the terminal falling speed \cite{Wilson2014}.   But a surprising difference is that the rate is not constant unless the filling height is very large: it actually {\it increases} as the hopper empties, ever faster near the end \cite{Wilson2014,OtherObservations}.

\begin{figure}[!ht]
\includegraphics[width=\columnwidth]{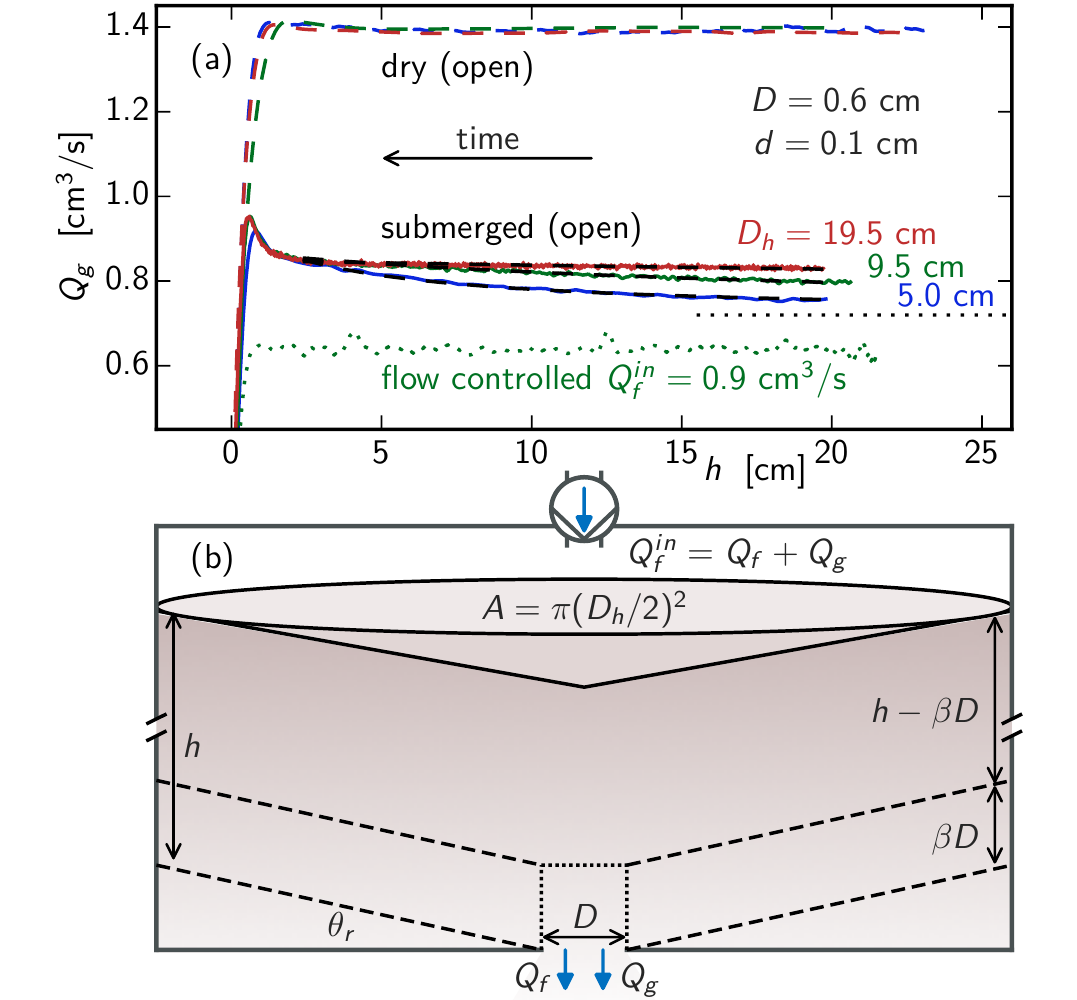}
\caption{(a) Granular discharge rate versus remaining height $h$ for 3 cases of dry (dashed) and submerged (solid) $d=0.1$~cm grains from a $D=0.6$~cm orifice in hoppers of diameters $D_h$, under several ``pressure controlled" conditions where the top of the hopper is open and also for one ``flow controlled" case that the top of the hopper is sealed and water is pumped in at a constant rate $Q_f^{in}$.  The black dashed curves represent fits to Eq.~(\ref{eq:simple}) over the range $h\ge \beta D$, as shown; these asymptote to the dotted line.  (b) Schematic illustration of the flow-control experiment, with defined quantities pertaining to hopper, grains, and fluid.}
\label{schema}
\end{figure}

This dramatic  ``surge" is important to understand because hopper flows are ubiquitous and are rarely in vacuum.  It is also important because interstitial fluid affects many granular phenomena, and is a crucial factor in suspensions, fluidization, and sedimentary transport \cite{GuazzelliMorris, FranklinShattuck}.  Furthermore, the mechanisms controlling hopper discharge rate are basic to the issue of clogging and the formation of stable arches \cite{ZuriguelPIP14, Thomas2015}.  To make progress our experimental approach is two-fold.  First, we measure discharge rate versus filling height with a more precise and automated apparatus, not just for submerged grains but also for dry grains in air (Fig.~\ref{schema}a).  Under these conditions, the flow of interstitial fluid is set by the granular discharge and can self-adjust as the hopper empties.  Second, we measure discharge where the interstitial water flow rate is fixed by pumping at a range of values.  We find that the granular discharge rate increases with fluid flow, but is constant in time with no surge.  Using these flow-control results as input, we model the surge by combining the hydrodynamic resistance of the grains in the hopper with pressure-control set by the grain exit speed.

Here we use technical quality monodisperse spherical glass beads (Potter Industries A-series).  The diameter distributions are nearly Gaussian, with mean and standard devation of either $d=0.498\pm0.048$~mm or $d=1.001\pm0.090$~mm (Retsch Technology Camsizer). The material density is $\rho_{glass} = 2.54\pm{0.01}~\mathrm{g/cm^3}$, found by sinking grains into water and measuring  the volume of displaced water versus the increase in mass.  Filtered tap water is used for the submerged cases.  Based on standard textbook values for density $\rho_f$ and viscosity $\eta$ at room temperature, the terminal velocity of the smaller (larger) beads is expected to be $v_t=7.5~(15.1)~\mathrm{cm/s}$ \cite{DragCoefficient}, in accord with visual observation.

For all experiments we use flat-bottomed cylindrical hoppers with concentric circular orifices (Fig.~\ref{schema}b and \cite{supp}).  The bottom plate consists of a polycarbonate disc with a 2.5~cm hole and a 5.1~cm depression to accommodate interchangeable aluminum discs 0.6~cm thick, each having an orifice of different diameter $D$.  The orifice consists of a cylindrical hole that extends 0.1~cm straight down from the top of the disc and then expands out in a 45~degree bevel cut to the bottom of the disc.  The hopper sidewalls consist of interchangeable polycarbonate tubes of desired inner diameter $D_h$, and 30~cm height, glued to a flange that bolts to the bottom plate.  This design makes it possible to vary the orifice diameter and hopper size reproducibly and independently.  The top of the hopper is either open, or is sealed off and connected to a gear pump (Cole Palmer 75210-50) to impose a desired volumetric flow rate $Q_f^{in}$ of water into the hopper with precision $\Delta Q_f^{in}=0.1~\mathrm{cm^3/s}$.

For all discharge measurements, the hopper hangs from a digital scale (Ohaus Valor 7000) with continuous readout to computer, whether in air or totally submerged in a tall aquarium (see \cite{supp}). The raw data is a set of mass-time pairs $m(t)$ with 1~g repeatability acquired at 10~Hz.  The mass of grains yet to be discharged is computed as $m_g(t) = [m(t)-m_{stop}]\rho_{glass}/(\rho_{glass}-\rho_{f})$, where the density factors account for buoyancy and where $m_{stop}=m(\infty)$ is the readout mass of the hopper and grains left inside when the flow stops at the end of the experiment.  The height of the grains yet to be discharged is calculated as $h(t)=m_g(t)/(\rho A)$, where $\rho=1.48\pm0.01~\mathrm{g/cm^3}$ is the density of the packing, determined in an auxiliary experiment where height $h(t)$ is obtained from camera images together with mass $m_g(t)$ from the scale. The volume fraction of the packing is thus $\phi = \rho/\rho_{glass} = 0.58\pm{0.04}$, consistent
with the loose packing results of Ref.~\cite{FarrellJFD10}.   The volumetric discharge rate of grains is $Q_g(t) =(-1/\rho_{glass})~\mathrm{d}m_g/\mathrm{d}t$, calculated by $2^{nd}$-degree polynomial fit over a window defined by Gaussian weighting with width $2\sigma=6$~s \cite{supp}.

Data for discharge rate $Q_g$ versus remaining height $h$ are plotted in Fig.~\ref{schema}a for the $D=0.6$~cm orifice and several different conditions: three different hopper diameters $D_h$ with an open top in air, and under water, and one hopper diameter under a fixed input water flow rate.  The quality of the data is far better than the pioneering observations, obtained by weighing the grains collected in a cup during a timed interval \cite{Wilson2014}.  With this improvement we now see that the surge depends on $D_h$, and that $\mathrm{d}Q_g/\mathrm{d}h$ is larger for smaller $D_h$.  We also discover a small surge for grains in air, for which we are aware of no precedents.  In accord with Beverloo, $Q_g$ appears constant and independent of $D_h$ until just before this terminal surge.  The third new feature in Fig.~\ref{schema}a is that under fixed fluid flow-control conditions, the granular discharge rate is constant; in particular, the surge effect is totally eliminated.  This indicates that the surge may be caused by interstitial fluid flow at a rate that increases as the hopper empties.

\begin{figure}[!th]
\includegraphics[width=\columnwidth]{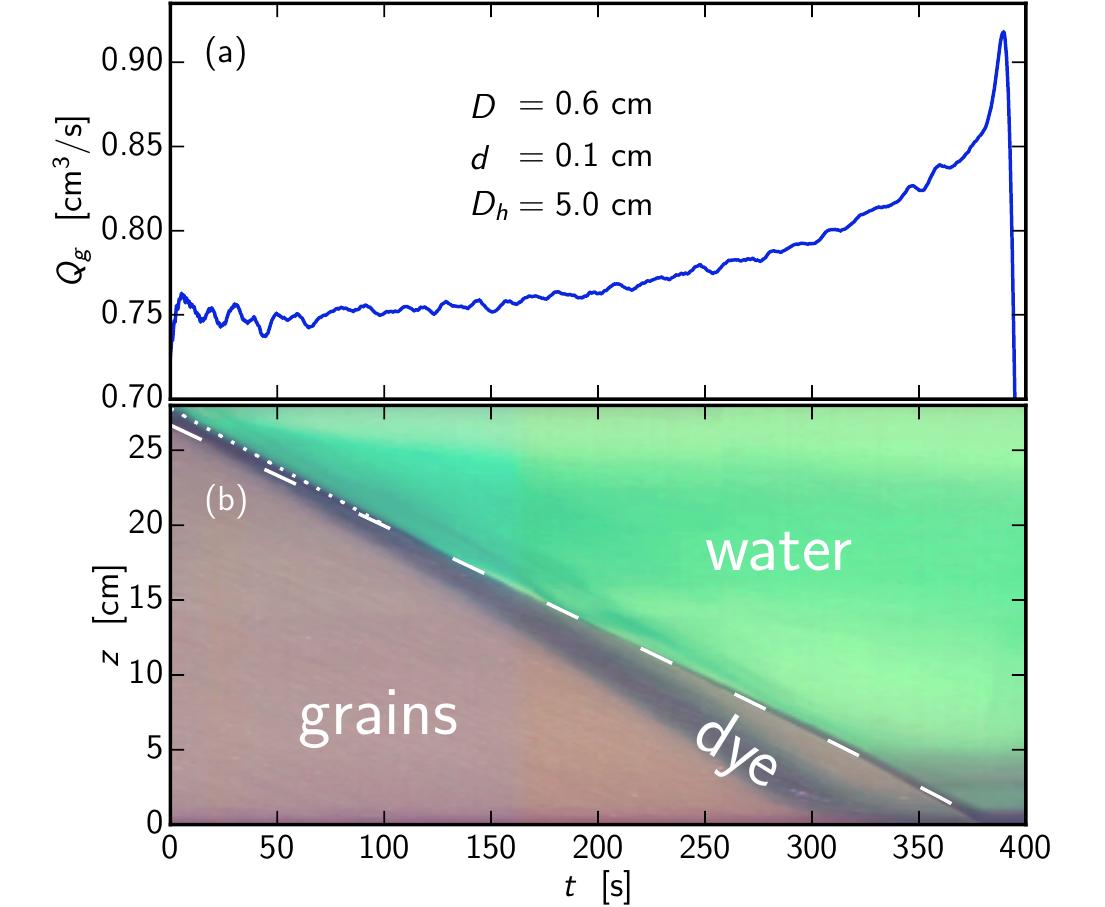}
\caption{(a) Granular discharge rate versus time, and (b) spacetime plot of the hopper from simultaneous video, for an open-top surge experiment. The spacetime plot is constructed by taking the vertical center line of the hopper side view from each time step $t$. Prior to commencing flow, a layer of concentrated dye was injected into the water $0-1$~cm above the packing.  The dashed white curve shows the top of the packing, as constructed from $Q_g(t)$ data, while the dotted line represents the propagation of the dye with flow rate $Q_f=0.62~\mathrm{cm^3/s}$, which holds for $Q_g=0.75~\mathrm{cm^3/s}$ according to the flow-control data in Fig.~\ref{flowcontrol}.  After $t=100$~s dye moves inside the packing ever more rapidly with time, as the surge increases.  The green color of the water is diluted dye from previous runs.}
\label{spacetime}
\end{figure}

To test this hypothesis qualitatively, we inject a layer a dye into the water just above the packing prior to the start of a surge experiment.  Fig.~\ref{spacetime}a shows discharge rate versus time, and underneath is a spacetime plot constructed from a simultaneous digital video recording \cite{supp}.  In it the grains appear brown, the dyed layer of water appears dark green, and the water above the packing appears light green due to small amounts of mixed dye.  The top edge of the packing is coincident with a dashed white curve constructed from $h(t)$.  With time, the discharge rate is seen to increase while the packing height decreases.  The dyed layer is seen to be above the packing at time zero, and to move down {\it into} the packing as time progresses.  And it moves faster in tandem with the increase in discharge rate.  Evidently, the act of granular discharge effectively creates a pumping effect whereby water flows down through the packing at a speed faster than the grains themselves. One might have guessed that the interstitial fluid would flow passively downward at the same speed as the grains, or perhaps slower, but in fact it moves faster.  This contrasts with prior work with sealed containers like an hourglass \cite{DuckerPT85, WuPRL93, VejeGM01, BerthoJFM02, MuitePOF04, HiltonPRE11}, where there is a backflow of air that volumetrically matches the downflow of grains.  Most recently, Ref.~\cite{HiltonPRE11} states that a common modeling approach for air backflow is an ``ad-hoc modification" of the Beverloo equation to include a pressure gradient opposing gravity.

\begin{figure}[!ht]
\includegraphics[width=\columnwidth]{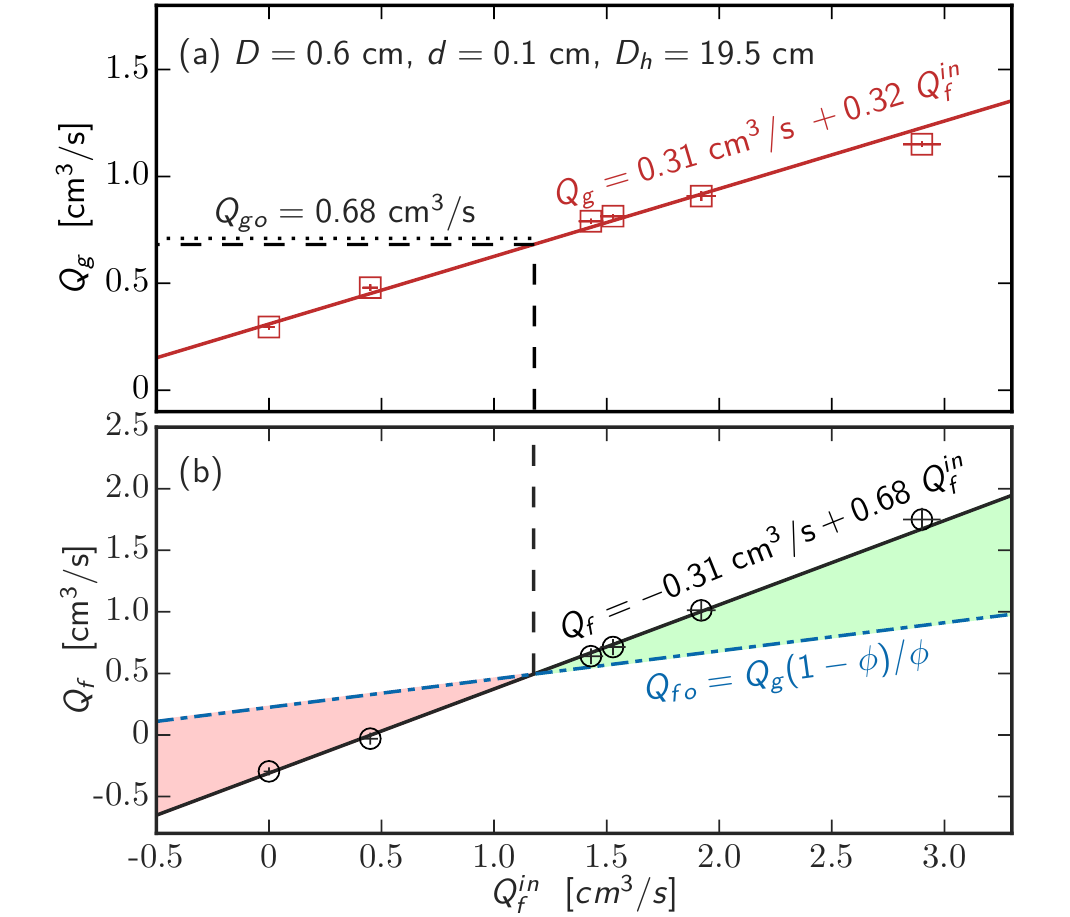}
\caption{Volumetric rate at which (a) grains and (b) water are discharged from a $D_h=19.5$~cm hopper, versus the rate at which water is pumped in at the top.  The grain and orifice diameters are the same as in Fig.~\ref{schema}a.   Results are well-described by line fits, as plotted.  The dash-dotted line in (b) represents the fluid output rate for passive flow along with the grains.  A positive difference $Q_f-Q_{fo}$ between actual and passive flow rates (green) will enhance the granular discharge, and vice-versa for a negative difference (red).  The crossing point specifies the reference rate $Q_{go}=0.68\pm0.05~\mathrm{cm^3/s}$ of granular discharge where the fluid flows passively with the grains (dashed line construction).  A consistent value of $Q'_{go}=0.72\pm0.01~\mathrm{cm^3/s}$ is found in Figs.~\ref{schema},\ref{modelfit} by fits of surge data to Eq.~(\ref{eq:simple}).}
\label{flowcontrol}
\end{figure}

To quantify the coupling between fluid and grains, we now perform a series of flow-control measurements for how the grain discharge rate $Q_g$ increases with fluid input pump rate $Q_f^{in}$.  Results are obtained from averaging over a time series (e.g.\ Fig.~\ref{schema}a) and are plotted in Fig.~\ref{flowcontrol}a.  These are well-described by a linear relation $Q_g = (0.310\pm{0.008})~\mathrm{cm^3/s} + (0.316\pm{0.007})~Q_f^{in}$.  This fit and the error estimates use weights based on uncertainty in pump rate as well as a 1\% uncertainty in $Q_g$.  By volume conservation, the rate at which fluid is pumped into the top of the hopper must equal the sum $Q_f^{in}=Q_g+Q_f$ of rates at which grains and fluid leave the orifice.  Results for $Q_g$ may thus be recast in terms of the fluid outflow rate $Q_f$, which, as shown in Fig.~\ref{flowcontrol}b, is also necessarily linear in $Q_f^{in}$.

To explain both the flow-control and the surge experiments, we begin by considering the excess or deficit of fluid flow with respect to the rate $Q_{fo}$ at which it flows passively with -- i.e. at the same speed as -- the grains in the hopper.  The simplest model is to assume a linear perturbation
\begin{equation}
	Q_g = Q_{go} + \alpha(Q_f-Q_{fo}),
\label{eq:excess}
\end{equation}
where $Q_{go}$ is the ``reference" grain discharge rate, when the fluid flow is passive, and $\alpha$ is a dimensionless proportionality constant.  By volume conservation, the passive fluid flow rate is given in terms of the grain discharge rate as $Q_{fo}=Q_{g}(1-\phi)/\phi$. The excess/deficit of fluid flow is indicated by the green/red shaded regions in Fig.~\ref{flowcontrol}b between the plotted lines for these two expressions. Their intersection graphically locates the passive reference state.  Inserting into Eq.~(\ref{eq:excess}) and rearranging gives a linear relation between the grain discharge rate and the rate at which fluid is pumped in:
\begin{equation} 
	Q_g =\left( \frac{ \phi }{ \alpha + \phi } \right) Q_{go}   + \left( \frac{ \alpha \phi}{\alpha + \phi} \right) Q_f^{in}.
\label{eq:pump}
\end{equation}
Comparing with the line fit in Fig.~\ref{flowcontrol}a gives the two unknowns as $\alpha=0.70\pm0.01$ and $Q_{go} = 0.68\pm0.05~\mathrm{cm^3/s}$, where the uncertainties reflects those in both the Fig.~\ref{flowcontrol}a fit and in $\phi$.  The grain and orifice diameters are the same as in Fig.~\ref{schema}a, where the discharge rates are all higher than this result for $Q_{go}$, as expected.  This same analysis of flow-control measurements is repeated for other grain and orifice sizes.  Similar to Ref.~\cite{Wilson2014}, results in \cite{supp} show that the reference discharge rate obeys a modified Beverloo equation,  $Q_{go}= C v_t d^2(D/d - k)^2$ where $C=0.4\pm0.1$ and $k=2.2\pm1.9$, and that the parameter $\alpha$ increases as $(D/d-k)^2$.  

Next we model the surge effect using Eq.~(\ref{eq:excess}) by considering the excess flow $Q_f-Q_{fo}=\Delta P/R$ of fluid through a porous granular medium of hydrodynamic resistance $R$ that is generated by a driving pressure $\Delta P$.  Since the excess fluid enters across the whole hopper area but exits through a much smaller orifice, the flow field is complex.  To simplify, we approximate the medium as two cylinders in series: the first has area $A=\pi(D_h/2)^2$ and height $h-\beta D$, while the second has area $\pi (D/2)^2$ and height $\beta D$, where $\beta$ is a dimensionless parameter that sets the height of the exit region where the permeation flow constricts and the grains dilate (see schematics in Fig.~\ref{schema}b and \cite{supp}).  The total hydrodynamic resistance is then
\begin{equation}
	R = \frac{\eta}{K d^2}   \left[    \frac{h-\beta D}{A} + \gamma \frac{\beta D}{\pi(D/2)^2}   \right],
\label{hydrores}
\end{equation}
where $K=(1-\phi)^3/(180\phi^2) = 0.00122$ (Kozeny-Carman equation for $\phi=0.58$ \cite{Beavers73, Dullien, SalBook}) and where $\gamma$ is an additional dimensionless parameter to account for the complex shape of the flow field and the increased permeability in the exit region.  These ingredients combine to predict the discharge rate versus the height $h$ of grains yet to exit as a sum of reference plus surge terms:
\begin{eqnarray}
Q_g &=& Q_{go} +  \alpha\frac{\Delta P K d^2 D / \eta}{(h-\beta D)D/A + 4\beta\gamma/\pi}, \label{eq:full}\\
        &=& Q_{go} +  \frac{a}{hD/A + b}. \label{eq:simple}
\end{eqnarray}
Eqs.~(\ref{eq:full}-\ref{eq:simple}) define $a$ and $b$ as convenient fitting parameters, and give their relation to $\beta$, $\gamma$, and $\Delta P$ as the underlying unknowns.  Eq.~(\ref{eq:simple}) also highlights the form of the surge with $h$ and hopper area $A$.  Namely, the surge vanishes (i.e.\ $Q_g=Q_{go}$) in the limit $h\rightarrow\infty$, where the hydrodynamic resistance is infinite and the interstitial fluid flows passively with the grains.  For smaller $h$ and larger $A$, the surge of $Q_g$ above $Q_{go}$ increases, just as seen in Fig.~\ref{schema}a, because the hydrodynamic resistance is smaller and the excess fluid flow is faster.

We now fit Eq.~(\ref{eq:simple}) to the three surge experiments in Fig.~\ref{schema}a, by adjusting $b$ separately for each data set but adjusting one value of $a$ and $Q'_{go}$ simultaneously for all.  The quality of the fits is good, and the values of the fitting parameters make sense:  First, the reference flow rate $Q'_{go}=0.72\pm0.01~\mathrm{cm^3/s}$ overlaps with the result from the flow-control experiments.  Second, the value of $a$ translates directly to a driving pressure of $\Delta P = a\eta/(\alpha K d^2 D)=5\pm3~\mathrm{Pa}$.  This is on the order of the Bernoulli pressure based on the single-grain terminal falling speed $v_t$, but is even closer to $\rho_f {v_s}^2/2$ where $v_s$ is the speed of the stream of discharged grains (see \cite{supp}).  Physically, the fluid pressure at the outlet is reduced due to some combination of grain dilation and fluid flow beneath the hopper, both of which are driven by gravity and hence might be expected to scale with $\rho_f{v_t}^2/2$.

\begin{figure}[!th]
\includegraphics[width=87mm]{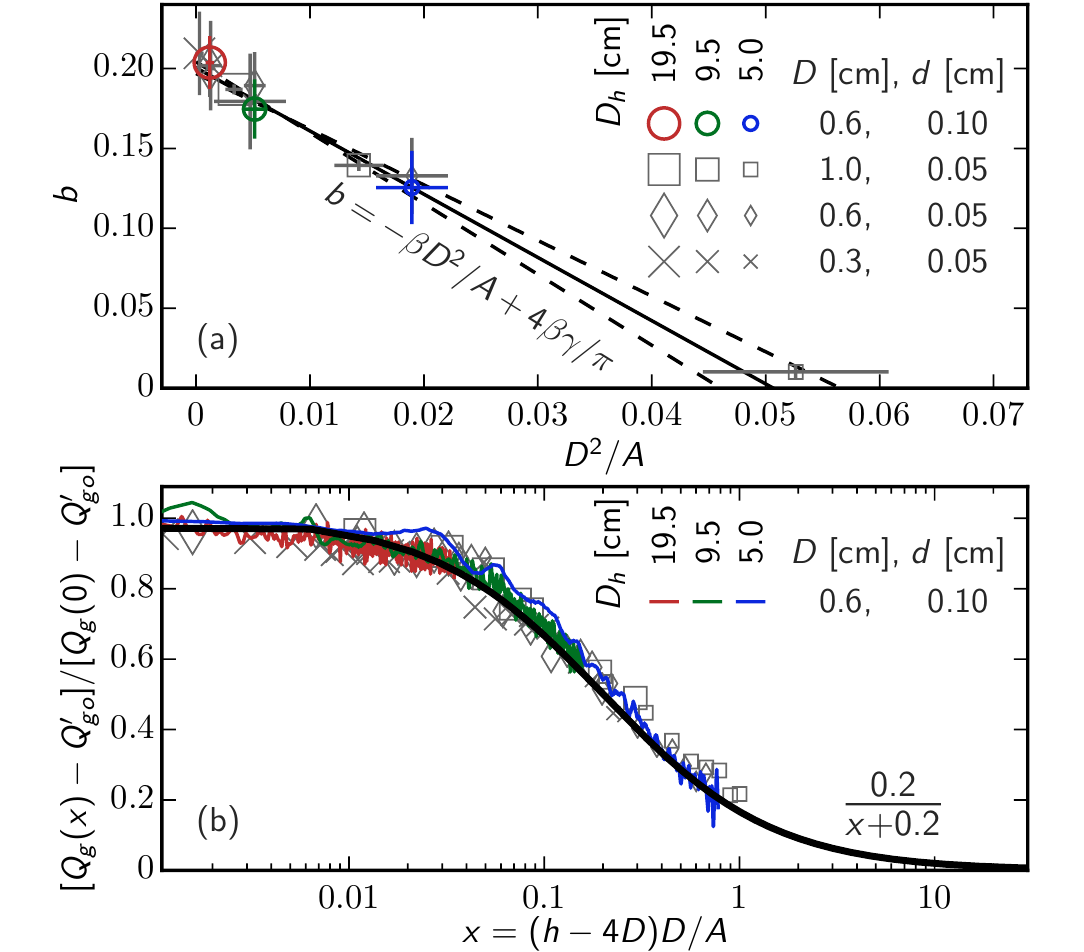}
\caption{(a) Parameter $b$ from fits of Eq.~(\ref{eq:simple}) to surge data for 4 combinations of particle and orifice sizes. Each labeled set contains 3 different experiments with hopper cross sectional areas $A=\pi(D_h/2)^2$; diameters $D_h=19.5, 9.5$ and $5.0$ cm decreasing left to right indicated by the symbol size. As predicted, $b$ decreases linearly with $D^2/A$.  The line fit has an intercept of $b_o=4\beta\gamma/\pi=0.200\pm0.004$ and a slope of $\beta = 4.0\pm 0.2$, giving $\gamma = 0.04\pm0.01$. (b) Plotting the data against $(h-\beta D)D/A$ collapses the data to a single curve as predicted by Eq.~(\ref{eq:full}) which asymptotes to $Q'_{go}$ at high packing fractions and to a value set by the flow rate at the constricting zone boundary $Q_g(h=\beta D)$. The gray symbols represent the same experiments in both figures. The colored datasets are the same as in Figure 1a.}
\label{modelfit}
\end{figure}

Next we repeat the surge experiments for the smaller beads and three hole and hopper sizes.  Fitting values for the parameter $b$ are plotted in Fig.~\ref{modelfit}a versus $D^2/A$.  As expected from Eqs.~(\ref{eq:full}-\ref{eq:simple}) the results decrease linearly with $D^2/A$, and do not depend on grain size.  The displayed line fit has slope $\beta=4.0\pm0.2$ and intercept $b_o=4\beta\gamma/\pi=0.200\pm0.004$; these combine to give $\gamma=0.04\pm0.01$.  The height of the region where the permeation flow becomes constricted is thus $\beta D=\mathcal O(D)$, as expected.  And the value of $\gamma$ is smaller than one, also as expected, because the grain density decreases in the exit region and because the true interstitial flow also enters through the side-walls of the imagined cylinder (see \cite{supp}).

As a final check we attempt to collapse all twelve surge data sets according to Eq.~(\ref{eq:full}), using only the fitting parameter $\beta=4$ and the reference discharge rates $Q'_{go}$.  In particular, we subtract $Q'_{go}$ and divide by the difference between $Q'_{go}$ and the rate at $h=\beta D$.  The scaled discharge rates then go between 0 and 1 as the $h$ decreases from infinity to $h=\beta D$.  And when plotted versus $x=(h-\beta D)D/A$, the data should all collapse to $b_o/(x+b_o)$.  As demonstrated in Fig.~\ref{modelfit}b, the scaled data all collapse beautifully to this form with the intercept $b_o=0.20$ found in Fig.~\ref{modelfit}a.  For $h>\beta D$ the surge is thus well-described and understood by the above model of grain-fluid coupling and the two-cylinder picture of the medium.  

The two-cylinder picture of the permeation flow ought to fail as $h$ decreases toward and below $\beta D$.  Nonetheless the fit to Eq.~(\ref{eq:simple}) holds well for all $h\ge \beta D$ as shown in the plots.  But for smaller $h$ there is an even more dramatic ``terminal'' surge where the discharge rate increases sharply, nearly independent of hopper diameter, as seen in Fig.~\ref{schema}a.  This is the only effect noticeable in Fig.~\ref{schema}a for dry grains in air.  To model the terminal surge would require a more sophisticated treatment of the shape of the packing, the permeation flow, and maybe even the motion of the grains.  And perhaps the driving pressure $\Delta P$ could no longer be treated as constant, but would increase in positive feedback with the surging discharge rate.  Further studies of the fluid pressure \cite{MorrisPRE10} and flow \cite{HiltonPRE11} fields would also be helpful.  In addition, it is possible that the terminal surge is an example of the ``faster-is-slower" effect \cite{AlonsoPRE12, PastorPRE15}.

In conclusion, with an improved apparatus we have uncovered the full phenomenology for the surge of granular discharge as a hopper empties.  We have demonstrated that this surprising effect originates in a permeation flow of interstitial fluid, which is pumped downward through the pack at a speed faster than the grains.  And we have quantitatively modeled this flow and its coupling to the granular discharge.  This is significant for establishing the baseline reference state, which ought to be the target for understanding the modified Beverloo law for the discharge rate of grains under water \cite{Wilson2014, supp}, as a very tall hopper where the interstitial fluid flows passively downwards at the same speed as the grains.  This raises a question about the usual Beverloo law for dry grains, where, both historically \cite{Beverloo, NeddermanSavage} and more recently \cite{MazaGM07, AguirrePRL2010, Hannah_GM10, JandaPRL12, RubioLargoPRL15}, interstitial air is neglected.  The baseline state of passive interstitial fluid flow may be even more important for understanding clogging \cite{ZuriguelPIP14}, where fluid must be squeezed out from between the grains forming a stable arch over the orifice, and where pumped fluid could break marginally stable arches.  Further in this regard, we now ask whether the fraction of microstates that precede a clog, as measured from the average discharge mass \cite{Thomas2015}, includes grain momenta degrees of freedom that are affected by the interstitial fluid.  

\begin{acknowledgments}
This work was supported by the Finnish Foundation's Post Doc Pool, Wihuri Foundation and Finnish Cultural Foundation (JK) and by the NSF through Grant No. DMR-1305199 (JK, DJD).
\end{acknowledgments}

%\bibliography{ReferencesHopper}

\end{document}

\def \myeqsimple {3}	

\begin{figure}[!ht]

\begin{center}

{\large\bf{Supplemental Figures for ``The Sands of Time Run Faster Near the End"}}\\
\mbox{}\\
Juha Koivisto and Douglas Durian\\
\emph{Department of Physics and Astronomy, University of Pennsylvania, Philadelphia, PA 19104, USA }
\end{center}

\vspace{1.0cm}
\includegraphics[width=5in]{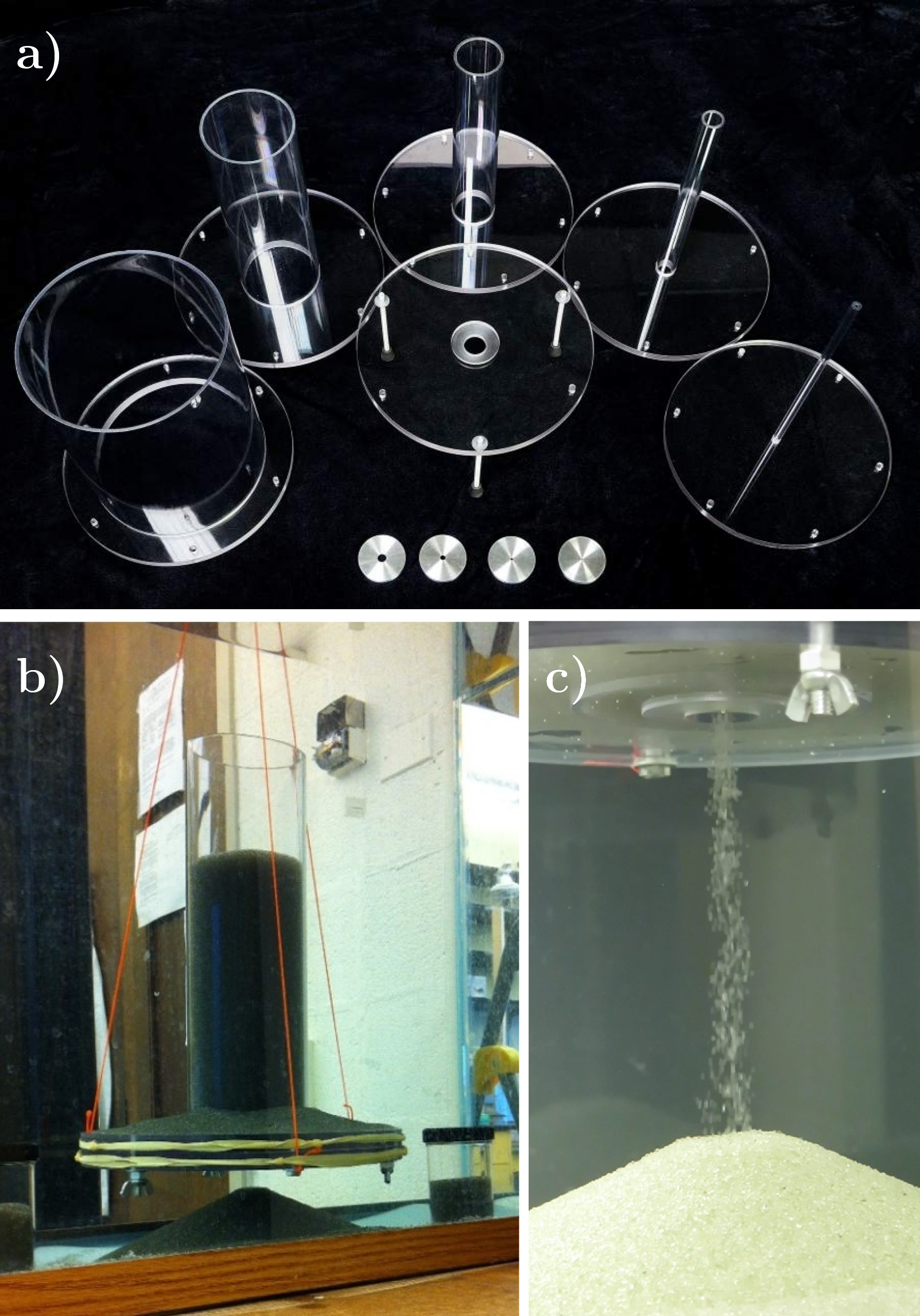}
\caption{(a) The measurement device consists of interchangeable polycarbonate tubes and 5.1~cm diameter aluminum disks with concentric orifices that can be fitted into the depression in the polycarbonate bottom plate shown at the center of the figure. 
(b) In the submerged case the hopper is completely under water in a fish tank, so that water can freely flow in at the top as grains exit at the bottom.  The orange strings connect the hopper to the digital scale.
(c) The $d=0.1$~cm grains flow in steady stream from the $D=0.6$~cm orifice, fully submerged under water.}
\end{figure}

% figure 2
\begin{figure}[ht]
\includegraphics[width=4.1in]{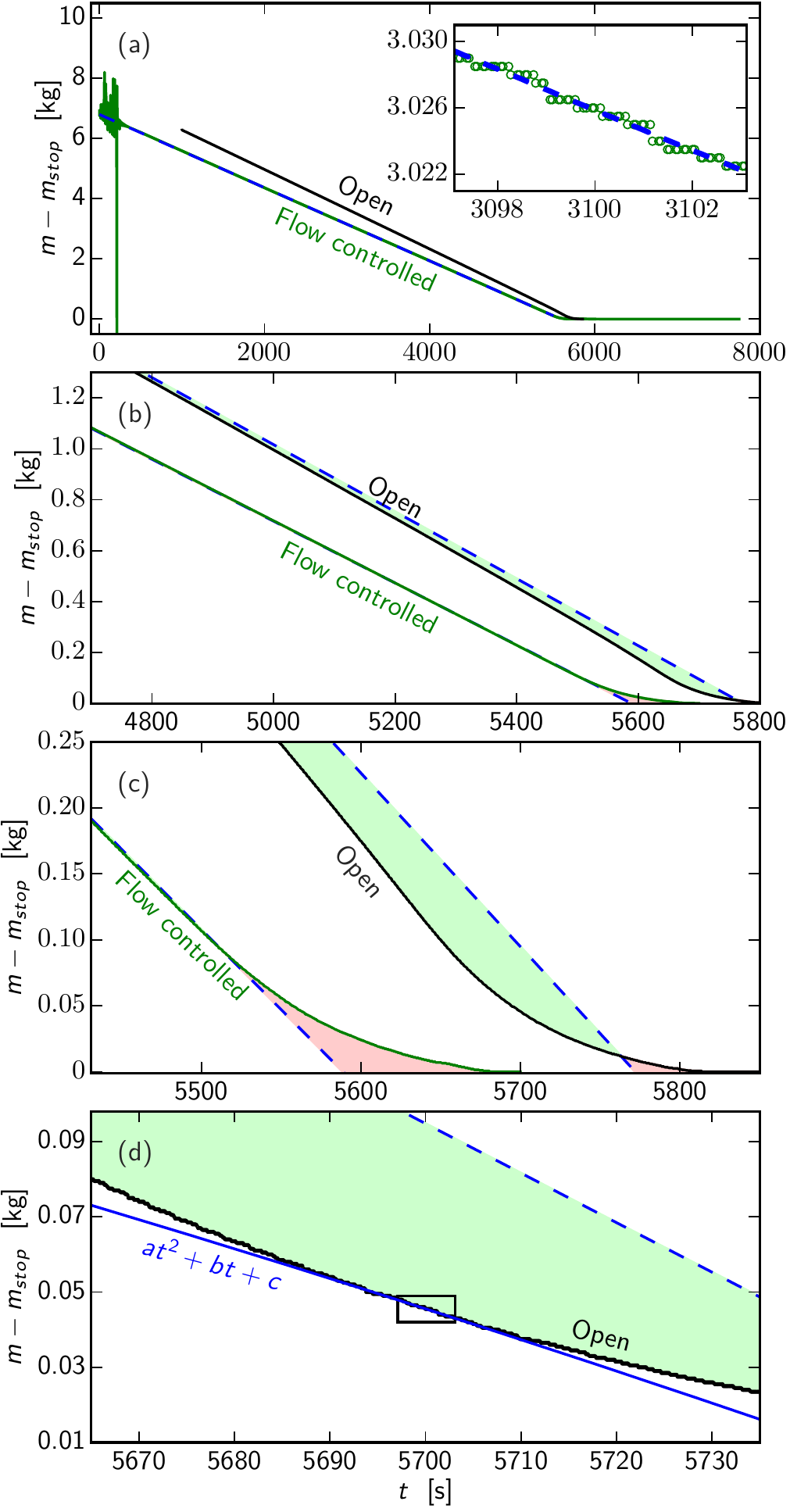}
\caption{(a) The raw data from the scale is plotted with the green curve (and circles in the inset). The final weight $m_{stop}$ at the end of the experiment when the grains stop flowing is subtracted from the data. At the beginning the experiment there are oscillations due to the starting procedure. The black curve is an open experiment with the same geometry. The dashed blue line is a linear fit from $t = 1000$~s to $t = 4000$~s that, \emph{in this figure only}, is considered the linear steady state. 
The inset is a magnification corresponding to differentiation window size with less than 1\% of the entire dataset.
Figures (b,c,d) are magnifications of Figure (a) with same curves and fits. The light green shaded areas highlight the increased flow rate in comparison with the linear fit. The increase in flow rate is only seen in the open case. The shaded red area at the end is the region where the hopper begins to run out of grains and the flow rate decreases. The black box in Figure (d) shows the Gaussian weighted differentiation window of $t_w = 2 \sigma = 6$ s used in the calculation of grain flow rate $Q_{go}$. The solid blue line in Figure (d) is the weighted $\mathrm{2^{nd}}$ degree polynomial fit $m - m_{stop} = a t^2 + b t + c$ to the data in the black box. The derivative $\mathrm{d}m\mathrm{/d}t$ is the first degree term $b$ in this fit.
}\label{fig:2_top}\label{fig:raw}
\end{figure}

% figure 3
\begin{figure}[ht]
\includegraphics[width=5in]{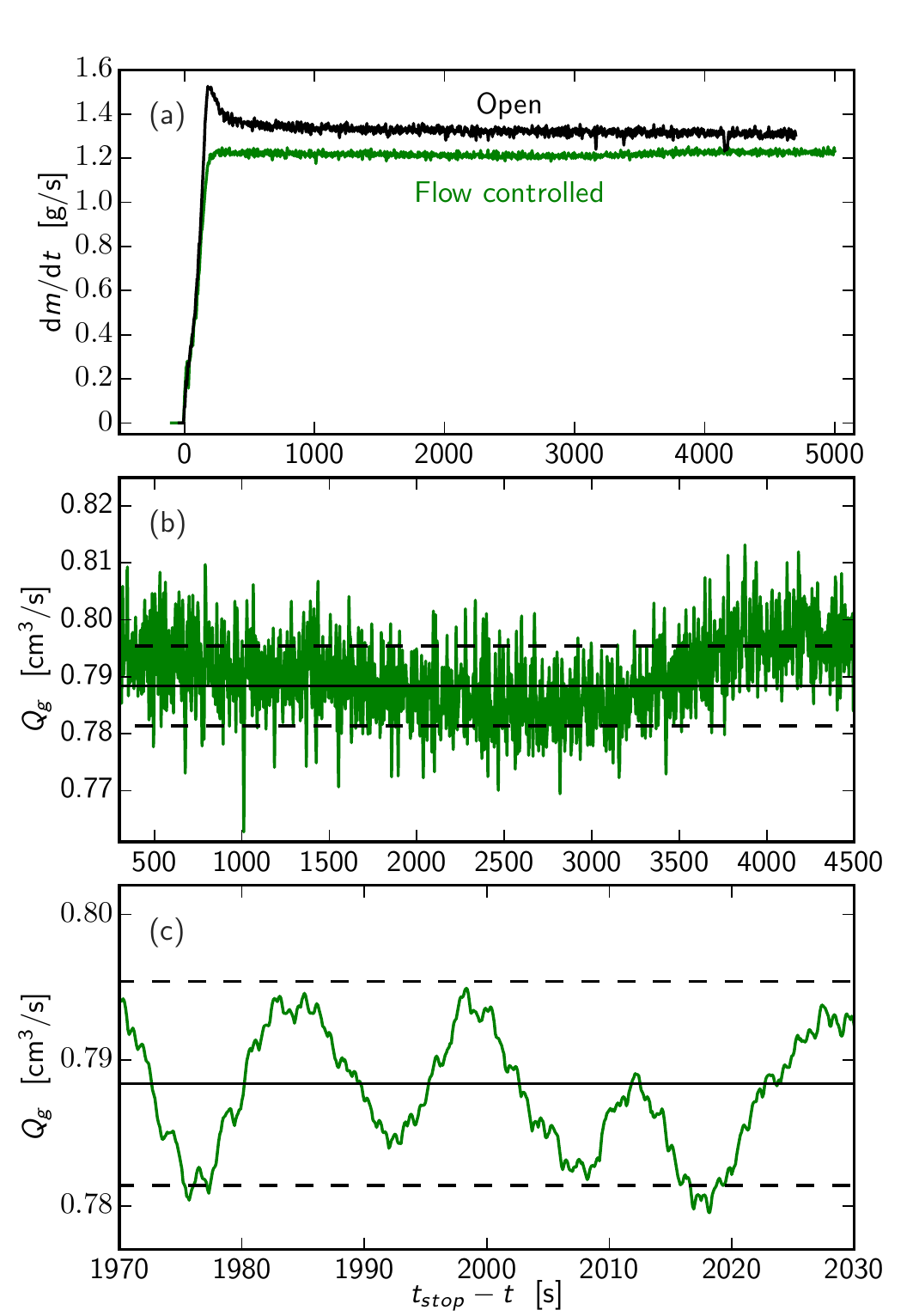}\\
\caption{(a) The mass flow rate $\mathrm{d}m/\mathrm{d}t$ from the windowed numerical derivative, for a flow controlled experiment (green curve) and an open experiment (black curve).  Both curves are from the same data as in Supplemental Figure~\ref{fig:raw}. The final ``stop"  time $t_{stop}$ is defined as when the flow stops and corresponds to the final weight $m_{stop}$. (b) The green data is the volume flow rate $Q_g = -(\rho_{glass} - \rho_{f})^{-1}\mathrm{d}m/\mathrm{d}t$, where the prefactor $(\rho_{glass} - \rho_{f})^{-1}$ takes the buoyancy of the water into account. 
The horizontal bars show the average flow rate (solid) and its error limits as one standard deviation from the mean. The data set is the same as in Figure~2a. c) The data in Figure~b is magnified. The oscillations in this scale are random walk with power spectrum that scales as Brownian noise exponent~2.
}\label{fig:2_bottom}\label{fig:dmdt}
\end{figure}

% figure 4
\begin{figure}[ht]
\includegraphics[width=5in]{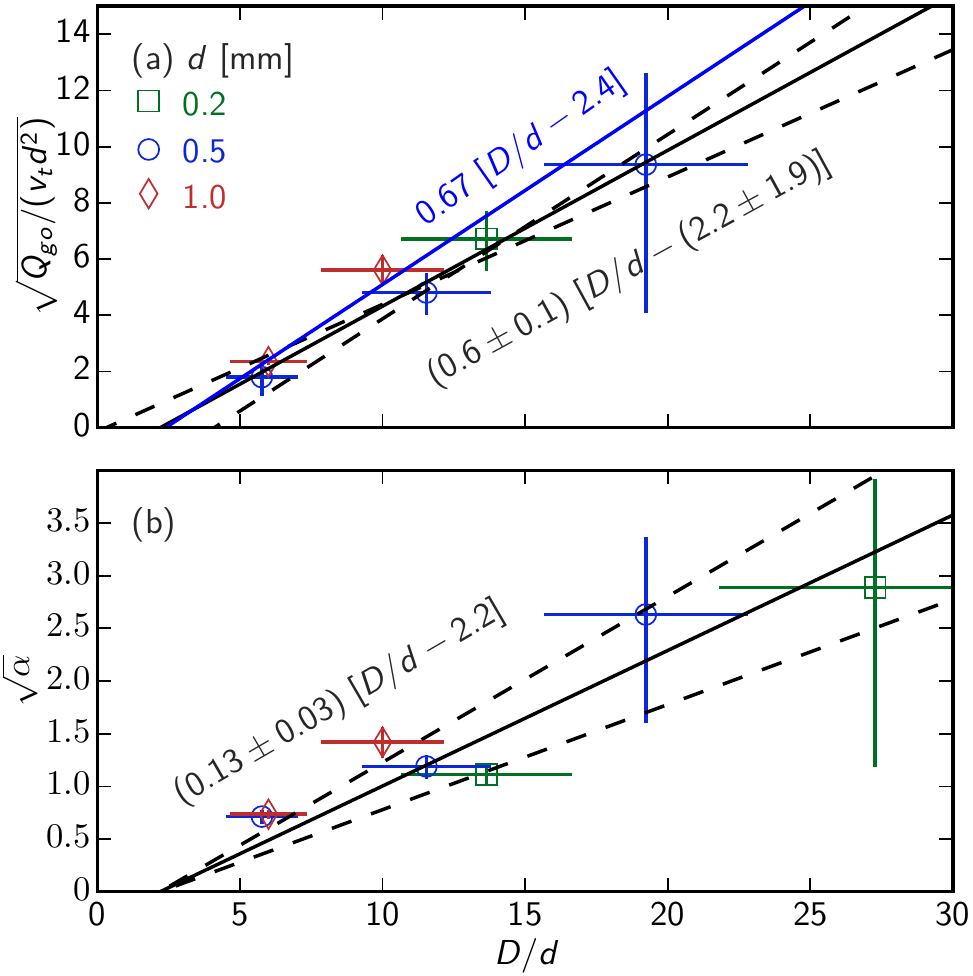}
\caption{(a) The reference flow rate $Q_{go}$, at which the fluid moves down passively at the speed as the grains, as obtained using the same method as in Figure~3 in the main article for different grain and orifice diameters. The black line fit depicts the modified Beverloo equation, $Q_{go}= C v_t d^2(D/d - k)^2$. The dimensionless fitting parameters are $C=0.6^2 = 0.4\pm0.1$ and $k = 2.2\pm1.9$, which are close to the reported values in Wilson {\it et al.} [Papers in Physics {\bf 6}, 060009 (2014)]:  $C_W = 0.45$ and $k_W=2.4\pm0.1$. Here the particle diameter and its uncertainty is taken from the manufacturer. The missing data point at $D/d = 27$ is off the scale in Figure (a) and disregarded as an outlier.
Figure (b) shows the excess fluid flow proportionality constant $\alpha$ defined by Eq.~(1) of the main text. The lines represent fits with the same cutoff constant ($k=2.2$) as in (a). In both figures, the line fits are weighted by the plotted statistical uncertainty error bars.}
\end{figure}

% figure 5
\vspace{10mm}
\begin{figure}[ht]
\includegraphics{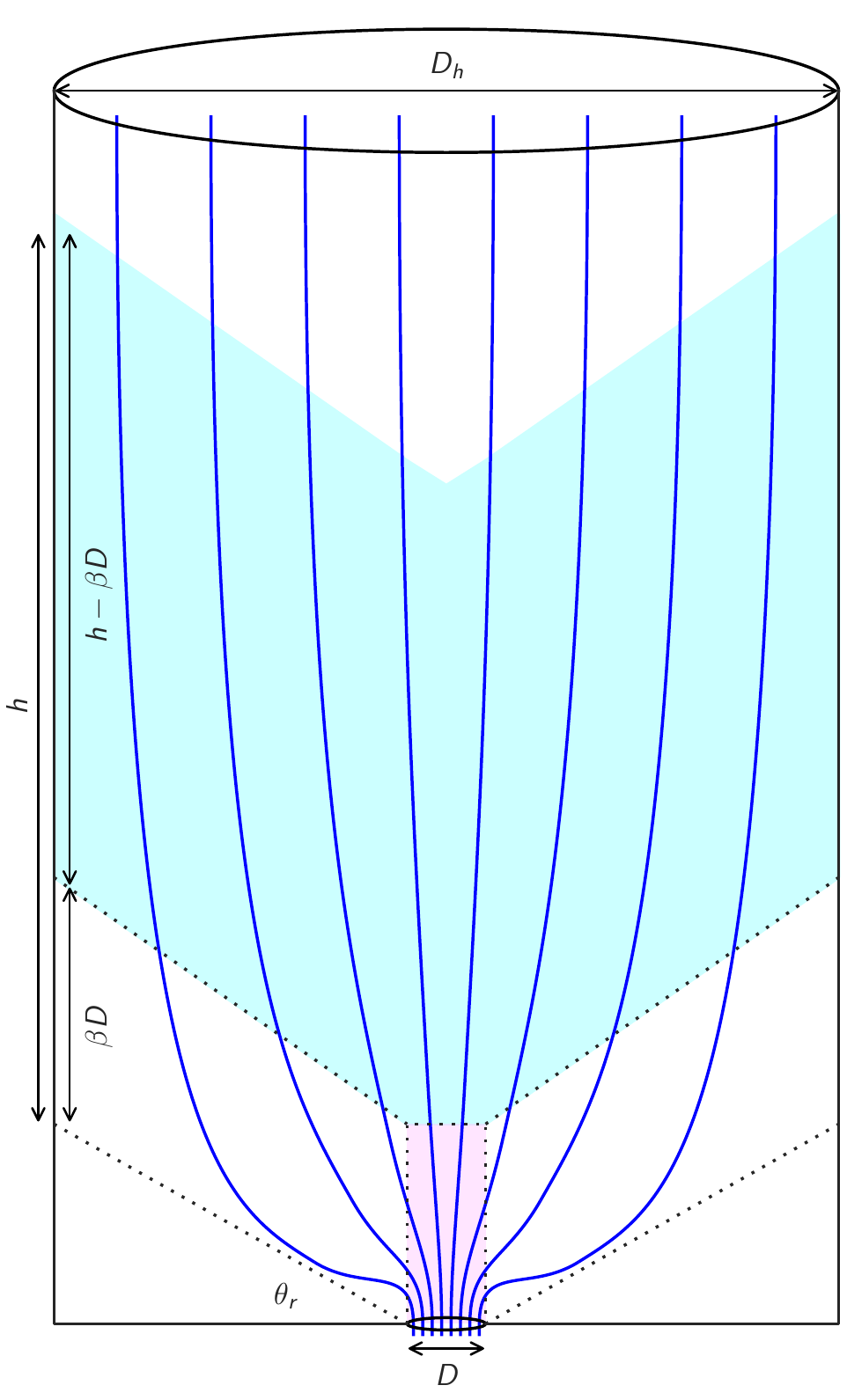}\\
\caption{Schematic illustration of the open experiment, similar to Fig.~1b of the main text.  The light blue and red shaded areas represent the two regions of the porous granular medium as defined by the hole diameter $D$, the parameter $\beta$, and the height $h$ of grains yet to be discharged.  The dark blue curves represent possible streamlines for the flow of fluid through the medium.  The angle of repose $\theta_r$ determines the set of grains that remain in the hopper after flow ceases, i.e.~when $h$ decreases to zero.}
\end{figure}

\FloatBarrier
\pagebreak

%figure 6
\begin{figure}[ht]
\includegraphics[width=4.5in]{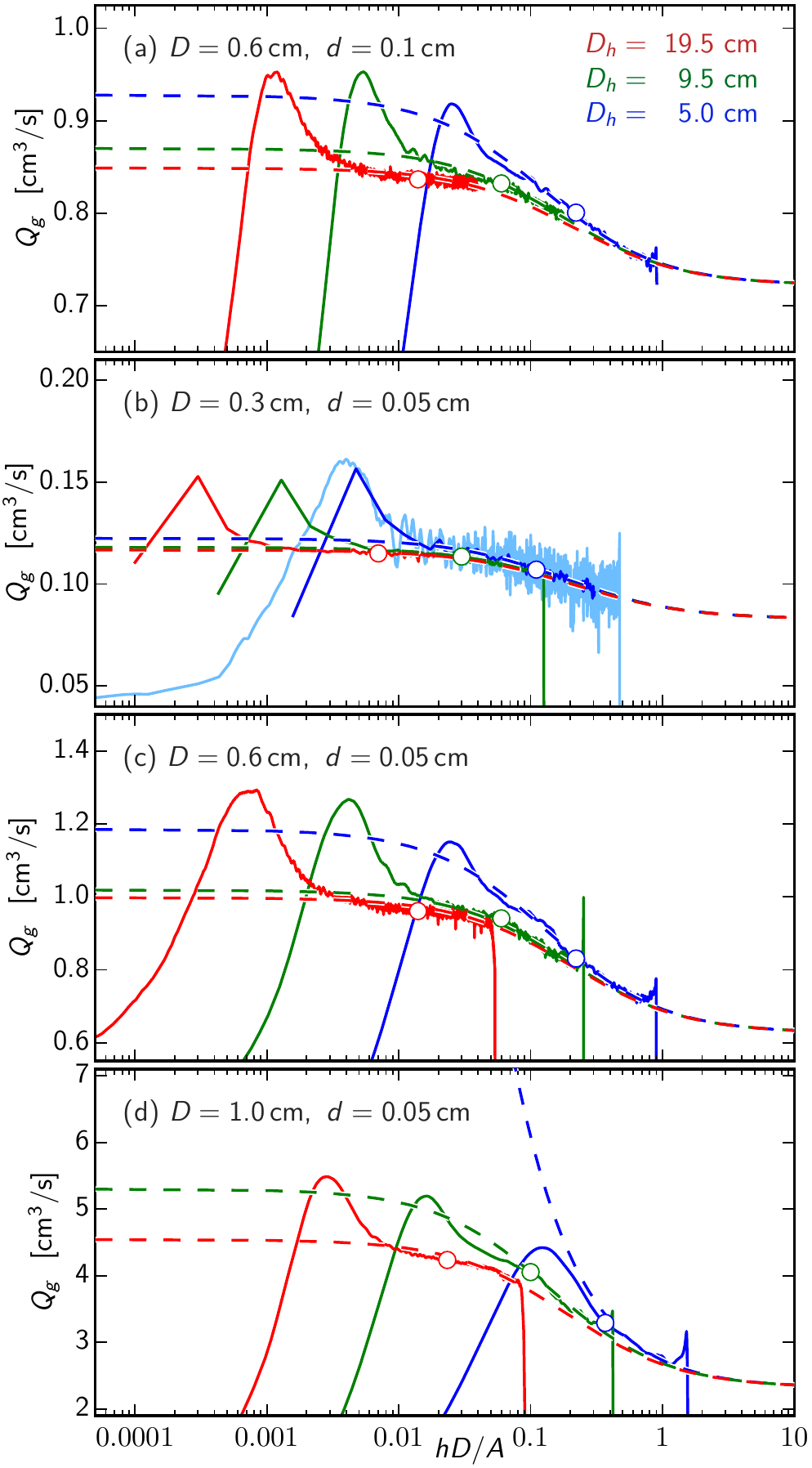}
\caption{Data and fits for open/surge discharge experiments with 12 combinations of grain, orifice and hopper diameters. The dashed curves represent Eq.~(5) from the main text, where $Q'_{go}$, $a$ and three values of $b$ are simultaneously fitted to the each set with same $d$ and $D$.  The white-filled circles indicate the lower limit for the fit region, $h \ge 7$~cm.   In (b) the $D = 0.3$ cm data is filtered with 2 mm windowed median. The original data is shown for $D_h=5.0$ cm case as a light blue swath. The relative noise level is higher in (b) as the flow rate is much smaller than in other cases.  All analyses for $D=0.3$~cm data use the filtered data.  Properly scaled, these data sets collapse together for $h\ge\beta D$ in Fig.~4b of the main text.}
\end{figure}

\FloatBarrier
\pagebreak

%figure 7
\begin{figure}[ht]
\includegraphics[width=5in]{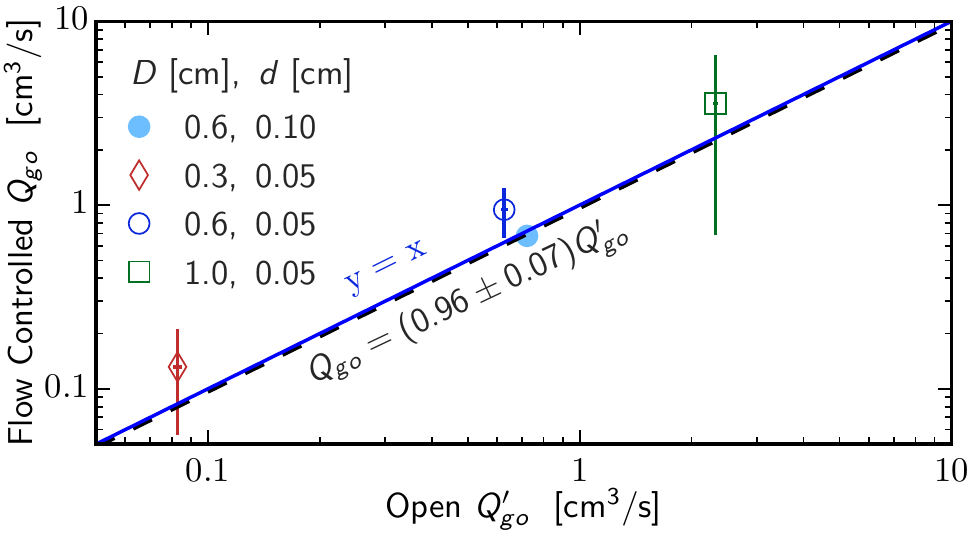}
\caption{Reference granular discharge rates, where the fluid flows passively with the grains,  from analysis of flow-controlled versus open/surge experiments.  Results fall near $y=x$, showing good agreement.  The dashed line is an error-weighted proportionality fit, $Q_{go} = (0.96\pm0.07)\;Q'_{go}$, consistent with $y=x$.}
\end{figure}

\FloatBarrier
\pagebreak

\begin{table}[ht]
\caption{Speeds and pressures for various diameter grains and orifices.  Here $v_t$ is the terminal speed of isolated grains, as computed from the drag coefficient [F.A.~Morrison, ``Data Correlation for Drag Coefficient for Sphere" (www.chem.mtu.edu/~fmorriso, accessed June 2012)]; the corresponding Reynolds numbers are 37 and 150 for 0.5 and 1.0~mm grains, respectively.  The speed $v_s$ of the stream of discharging grains is measured by video with an accuracy of 5~cm/s.  These speeds translate to characteristic Bernoulli and viscous pressure scales, as tabulated.  For comparison, the fitting parameter $a$ defined by Eqs.~(4,5) of the main text gives the actual pressures $\Delta P$, in the third column, that drive the permeation flow.  Note that $\Delta P$ corresponds most closely to the Bernoulli pressure $\rho_f {v_s}^2/2$, indicating that  fluid flow excited under the hopper by the falling stream of grains is the source of pressure that pumps water through the grains and that -consequently- causes the surge effect.} 
\vspace{0.5cm}

\begin{tabular}{|c c rl | rl rl c |rl rl c|c c c c}
\hline
$~d~\mathrm{[cm]}~$  & $~D~\mathrm{[cm]}~$ & \multicolumn{2}{c|}{$~\Delta P~\mathrm{[Pa]}~$} & \multicolumn{2}{c}{$~v_{t}~\mathrm{[cm/s]}~$}  & \multicolumn{2}{c}{$~\rho_fv_t^2/2~\mathrm{[Pa]}$~} &~$\eta v_t/d$~[Pa]~&  \multicolumn{2}{c}{$~v_{s}~\mathrm{[cm/s]}~$}  &  \multicolumn{2}{c}{~$\rho_fv_s^2/2~\mathrm{[Pa]}$}~& ~$\eta v_s/d$~[Pa]~\\ 
\hline
% d         D      \Delta P                                      v_t                                                                  r vt^2/2     			           eta v_t/d          v_s                                                    %                   r vs^2/2                        eta v_s/d
 0.10  & 0.6 &\hspace{0.05cm}5 & $\pm$ 3  &  \hspace{0.1cm}15.1 & $\pm$ 0.8     &  \hspace{0.4cm}11  & $\pm$ 1     &   0.15     & \hspace{0.2cm}25& $\pm$ 5   &  \hspace{0.35cm}30 & $\pm$ 20  &      0.25          \\
 0.05  & 0.3 &                       15  & $\pm$ 4    & 									 7.5  & $\pm$ 0.4     &  									2.8 & $\pm$ 0.3   &   0.15     &                       15& $\pm$ 5  &                        11 & $\pm$ 8   &      0.30          \\ 
 0.05  & 0.6 &                       29  & $\pm$ 7    &  									 7.5  & $\pm$ 0.4     &                      2.8 & $\pm$ 0.3   &   0.15     &                       30& $\pm$ 5   &                        50 & $\pm$ 20  &      0.60          \\
 0.05  & 1.0 &                       20  & $\pm$ 10  &  									7.5  & $\pm$ 0.4     &                       2.8 & $\pm$ 0.3   &   0.15     &                       30& $\pm$ 5   &                        50 & $\pm$ 20  &      0.60          \\
\hline
\end{tabular}
\end{table}